# Frequency of Correctness versus Average Polynomial Time and Generalized Juntas[*]

Gábor Erdélyi[†]    Lane A. Hemaspaandra[‡]    Jörg Rothe[†]    Holger Spakowski[§]

June 16, 2008

**Abstract**

We prove that every distributional problem solvable in polynomial time on the average with respect to the uniform distribution has a frequently self-knowingly correct polynomial-time algorithm. We also study some features of probability weight of correctness with respect to generalizations of Procaccia and Rosenschein's junta distributions [PR07b].

[*]Also appears as URCS TR-2008-934. Supported in part by NSF grant CCF-0426761, DFG grants RO 1202/{9-1, 9-3, 11-1, 12-1}, the European Science Foundation's EUROCORES program LogICCC, the Alexander von Humboldt Foundation's TransCoop program, a Friedrich Wilhelm Bessel Research Award, and the "CompView" Global COE, which supported Lane A. Hemaspaandra's visit to the Tokyo Institute of Technology. Some of the results of this paper were presented at the *First International Workshop on Computational Social Choice* and at the *Sixteenth International Symposium on Fundamentals of Computation Theory*, August 2007 [EHRS07a].

This technical report in part updates the earlier technical report that appeared in March 2007 as URCS-TR-914 and as arxiv.org report cs.0703097. However, some parts of this version point to discussions in that earlier version of issues that are covered there in more detail. On the other hand, that earlier report's entire discussion of the weighted lobbying problem should be ignored, as Tuomas Sandholm has kindly pointed out to us that the weighted lobbying problem can be captured as a special case of reverse combinatorial auctions, and thus that our approximation result can be easily obtained from the earlier and very elegant work on approximation for reverse combinatorial auctions by Sandholm, Suri, Gilpin, and Levine [SSGL02]. We are very grateful to him for pointing this out. We mention in passing that, following a direction insightfully suggested by Tuomas, there is an easy-to-see polynomial-time reduction from reverse combinatorial auctions to the weighted lobbying problem, and in that reduction the size of the input is at most doubled, and so lower bounds on approximation for the former are inherited, with only a very slight degradation, by the latter.

[†]Institut für Informatik, Heinrich-Heine-Universität Düsseldorf, 40225 Düsseldorf, Germany. URLs: ccc.cs.uni-duesseldorf.de/{∼ erdelyi,∼ rothe}.

[‡]Department of Computer Science, University of Rochester, Rochester, NY 14627, USA. URL: www.cs.rochester.edu/u/lane. Work done in part while visiting Heinrich-Heine-Universität Düsseldorf and the Tokyo Institute of Technology.

[§]Department of Mathematics & Applied Mathematics, University of Cape Town, Rondebosch 7701, South Africa. URL: www.mth.uct.ac.za/∼ webpages/spakowski. Work done in part while at Heinrich-Heine-Universität Düsseldorf.

# 1 Introduction

Preference aggregation and election systems have been studied for centuries in social choice theory, political science, and economics, see, e.g., Black [Bla58] and McLean and Urken [MU95]. Recently, these topics have become the focus of attention in various areas of computer science as well, such as artificial intelligence (especially with regard to distributed AI in multiagent settings), systems (e.g., for spam filtering), and computational complexity. Faliszewski et al. [FHHR] provides a survey of some recent progress in complexity-related aspects of elections.

This paper's work, while not directly about elections, is motivated by models and notions from two papers that are from the study of elections, namely, the work of Homan and Hemaspaandra on greedy algorithms for Dodgson elections [HH] and the work of Procaccia and Rosenschein on the relationship between junta distributions and manipulation of elections [PR07b].

The Dodgson winner problem was shown NP-hard by Bartholdi, Tovey, and Trick [BTT89]. Hemaspaandra, Hemaspaandra, and Rothe [HHR97] optimally improved this result by showing that the Dodgson winner problem is complete for $P_{\|}^{NP}$, the class of problems solvable via parallel access to NP. Since these hardness results are in the worst-case complexity model, it is natural to wonder if one at least can find a heuristic algorithm solving the problem efficiently for "most of the inputs occurring in practice." Homan and Hemaspaandra ([HH], see also the closely related work of McCabe-Dansted, Pritchard, and Slinko [MPS]; [HH] discusses in detail the similarities and contrasts between the two papers' work) proposed a heuristic, called Greedy-Winner, for finding Dodgson winners. They proved that if the number of voters greatly exceeds the number of candidates (which in many real-world cases is a very plausible assumption), then their heuristic is a *frequently self-knowingly correct algorithm*, a notion they introduced to formally capture a strong notion of the property of "guaranteed success frequency" [HH]. We study this notion in relation with average-case complexity.

We also investigate Procaccia and Rosenschein's notion of deterministic heuristic polynomial time for their so-called junta distributions, a notion they introduced in their study of the "average-case complexity[1] of manipulating elections" [PR07b]. We show that under the junta definition, when stripped to its basic three properties, every NP-hard set is $\leq_m^p$-reducible to a set in deterministic heuristic polynomial time relative to some junta distribution and we also show a very broad class of sets (including many NP-complete sets) to be in deterministic heuristic polynomial time relative to some junta distribution.

This paper is organized as follows. In Section 2, we show that every problem solvable in average-case polynomial time with respect to the uniform distribution has a frequently self-knowingly correct polynomial-time algorithm. In Section 3, we study Procaccia and Rosenschein's junta distributions. The appendices present the heuristic Greedy-Score on which Greedy-Winner is based and the notion of frequently self-knowingly correct algorithm [HH] as well as some needed technical definitions from average-case complexity theory [Lev86, Imp95, Gol97, Wan97].

---

[1] To avoid any confusion, it is important to keep in mind (see also [Tre02]) that the "average-case complexity" results of [PR07b] are not average-case complexity results in the sense of being about some sort of averaging of running times, but rather they use the term to apply to results about frequency of correctness—or, to be more precise, probability weight of correctness (which is the type of result that both their paper and the paper of Homan and Hemaspaandra obtain). In this paper, for clarity, we will not use the term average-case in that fashion (except in quotation marks).

## 2 Frequency of Correctness versus Average-Case Polynomial Time

### 2.1 A Motivation: How to Find Dodgson Winners Frequently

An election $(C, V)$ is given by a set $C$ of candidates and a set $V$ of voters, where each vote is specified by a preference order on all candidates and the underlying preference relation is strict (i.e., irreflexive and antisymmetric), transitive, and complete. A Condorcet winner of an election is a candidate $i$ such that for each candidate $j \neq i$, a strict majority of the voters prefer $i$ to $j$. Not all elections have a Condorcet winner, but when a Condorcet winner exists, he or she is unique. In 1876, Dodgson [Dod76] proposed an election system that is based on a combinatorial optimization problem: An election is won by those candidates who are "closest" to being a Condorcet winner. More precisely, given a Dodgson election $(C, V)$, every candidate $c$ in $C$ is assigned a score, denoted by DodgsonScore$(C, V, c)$, which gives the smallest number of sequential exchanges of adjacent preferences in the voters' preference orders needed to make $c$ a Condorcet winner with respect to the resulting preference orders. Whoever has the lowest Dodgson score wins.

The problem Dodgson-Winner is defined as follows: Given an election $(C, V)$ and a designated candidate $c$ in $C$, is $c$ a Dodgson winner in $(C, V)$? (The search version of this decision problem can easily be stated.) As mentioned earlier, Hemaspaandra, Hemaspaandra, and Rothe [HHR97] have shown that this problem is $P_\|^{NP}$-complete.

It certainly is not desirable to have an election system whose winner problem is hard, as only systems that can be evaluated efficiently are actually used in practice. Fortunately, there are a number of positive results on Dodgson elections and related systems as well. In particular, Bartholdi, Tovey, and Trick [BTT89] proved that for elections with a bounded number of candidates or voters Dodgson winners are asymptotically easy to determine. Fishburn [Fis77] proposed a "homogeneous" variant of Dodgson elections that Rothe, Spakowski, and Vogel [RSV03] proved to have a polynomial-time winner problem. McCabe-Dansted, Pritchard, and Slinko [MPS] proposed a scheme (called Dodgson Quick) that approximates Dodgson's rule with an exponentially fast convergence. Homan and Hemaspaandra ([HH], see also McCabe-Dansted, Pritchard, and Slinko [MPS]) proposed a greedy heuristic that finds Dodgson winners with a "guaranteed high frequency of success." To capture a strengthened version of this property formally, they introduced the notion of a "frequently self-knowingly correct algorithm" (see Appendix A for the formal definition and for their heuristic Greedy-Score).

## 2.2 On AvgP and Frequently Self-Knowingly Correct Algorithms

Our main result in this section relates polynomial-time benign algorithm schemes (see Definition B.2 in Appendix B) to frequently self-knowingly correct algorithms (see Definition A.1 in Appendix A). We show that every distributional problem that has a polynomial-time benign algorithm scheme with respect to the uniform distribution must also have a frequently self-knowingly correct polynomial-time algorithm. It follows that all uniformly distributed AvgP problems have a frequently self-knowingly correct polynomial-time algorithm.

**Theorem 2.1** *Suppose that $\mathcal{A}(x, \delta)$ is a polynomial-time benign algorithm scheme for a distributional problem $f$ on $\hat{\mu}$ (the standard uniform distribution, see Appendix B). Then there is a frequently self-knowingly correct polynomial-time algorithm $\mathcal{A}'$ for $f$.*

**Proof.** Let $\delta(n) = 1/(n+1)^3$. Define algorithm $\mathcal{A}'$ as follows:

1. On input $x \in \Sigma^*$, simulate $\mathcal{A}(x, \delta(|x|))$.

2. If $\mathcal{A}(x, \delta(|x|))$ outputs ?, then output (*anything*, "*maybe*").

3. If $\mathcal{A}(x, \delta(|x|))$ outputs $y \in T$, where $y \neq ?$, then output ($y$, "*definitely*").

By Definition B.2 (which is given in Appendix B), algorithm $\mathcal{A}'$ runs in polynomial time. It remains to show that $\mathcal{A}'$ is frequently self-knowingly correct.

Fix an arbitrary $n \in \mathbb{N}^+$. Now, we must be careful regarding the fact that Impagliazzo's definition of benign algorithm schemes and its "$\delta$" guarantees are all with regard to drawing not over inputs of a given length (which is what we wish to consider) but rather regarding drawing from inputs *up to and including* a given length. Thus, there is some danger that even if a benign algorithm performs well when its length parameter is $n$ (meaning related to strings of length up to and including $n$), that such a "good" error frequency might be due not to goodness at length $n$ but rather to goodness at lengths $n-1$, $n-2$, and so on. However, if one looks carefully at the relative weights of the different lengths this is at most a quadratically weighted effect (that is, the distribution's probability weight at length $n$ is just quadratically less than the weight summed over all lengths less than $n$), and so our choice of $\delta(n) = 1/(n+1)^3$ is enough to overcome this.

Let us now handle that rigorously. Recall that $n$ is fixed and arbitrary. Let us set the constant (for fixed $n$) $\delta'$ to be $1/(n+1)^3$. So, clearly

$$\text{Prob}_{\hat{\mu}_{\leq n}}[\mathcal{A}(x, \delta') = ?] =$$
$$\frac{\sum_{i=1}^{n-1} 1/(i(i+1))}{\sum_{i=1}^{n} 1/(i(i+1))} \text{Prob}_{\hat{\mu}_{\leq n-1}}[\mathcal{A}(x, \delta') = ?] + \frac{1/(n(n+1))}{\sum_{i=1}^{n} 1/(i(i+1))} \text{Prob}_{\hat{\mu}_n}[\mathcal{A}(x, \delta') = ?].$$

Since $\mathcal{A}$ is a benign algorithm scheme, $\text{Prob}_{\hat{\mu}_{\leq n}}[\mathcal{A}(x, \delta') = ?] \leq \delta'$. So, combining this and the above equality, and solving for $\text{Prob}_{\hat{\mu}_n}[\mathcal{A}(x, \delta') = ?]$, we have

$$\text{Prob}_{\hat{\mu}_n}[\mathcal{A}(x, \delta') = ?] \leq$$
$$\frac{\sum_{i=1}^{n} 1/(i(i+1))}{1/(n(n+1))} \left( \delta' - \frac{\sum_{i=1}^{n-1} 1/(i(i+1))}{\sum_{i=1}^{n} 1/(i(i+1))} \text{Prob}_{\hat{\mu}_{\leq n-1}}[\mathcal{A}(x, \delta') = ?] \right).$$

And so, clearly, $\text{Prob}_{\hat{\mu}_n}[\mathcal{A}(x, \delta') = ?] \leq n(n+1)\delta' = n(n+1)/(n+1)^3$. So

$$\lim_{n \to \infty} \frac{\|\{x \in \Sigma^n \mid \mathcal{A}'(x) \in T \times \{\text{"maybe"}\}\}\|}{\|\Sigma^n\|} = 0,$$

which completes the proof. ❑

**Corollary 2.2** *Every distributional problem that under the standard uniform distribution is in* AvgP *has a frequently self-knowingly correct polynomial-time algorithm.*

**Proof.** Impagliazzo proved that any distributional problem on input ensemble $\mu_n$ is in AvgP if and only if it has a polynomial-time benign algorithm scheme; see Proposition 2 in [Imp95]. The claim now follows from Theorem 2.1. ❑

It is easy to see that the converse implication of that in Corollary 2.2 is not true.

**Proposition 2.3** *There exist (distributional) problems with a frequently self-knowingly correct polynomial-time algorithm that are not in* AvgP *under the standard uniform distribution.*

**Proof.** For instance, one can define a problem that consists only of strings in $\{0\}^*$ encoding the halting problem. This problem is clearly not in AvgP, yet it is frequently self-knowingly correct. ❑

## 3   Frequency of Correctness versus Basic Junta Distributions: A Basic Junta Distribution for SAT

Procaccia and Rosenschein [PR07b] introduced "junta distributions" in their study of NP-hard manipulation problems for elections. The goal of a junta is to be such a hard distribution (that is, to focus so much weight on hard instances) that, loosely put, if a problem is easy relative to a junta then it will be easy relative to any reasonable distribution (such as the uniform distribution). This is a goal, not (currently) a theorem; Procaccia and Rosenschein [PR07b] do not formally establish this, but rather seek to give a junta definition that might satisfy this. Their paper in effect encourages others to weigh in and study the suitability of the notion of a junta and the notion built on top of it, heuristic polynomial time.

Regarding Procaccia and Rosenschein's notion of juntas, they state three "basic" conditions for a junta, and then give two additional ones that are tailored specifically to the needs of NP-hard voting manipulation problems. They state their hope that their scheme will extend more generally, using the three basic conditions and potentially additional conditions, to other mechanism problems. One might naturally wonder whether their junta/heuristic polynomial-time/susceptibility approach applies more generally to studying the probability weight of correctness for NP-hard problems, since their framework in effect (aside from the two "additional" junta conditions just about voting manipulation) is a general one relating problems to probability weight of correctness. We first carefully note that in asking this we are taking their notion beyond the realm for which it was explicitly designed, and so we do not claim to be refuting any claim of their paper. What we will do, however, is show that the three basic conditions for a junta are sufficiently weak that one can construct a junta relative to which the standard NP-complete problem SAT—and a similar attack can be carried out on a wide range of natural NP-complete problems—has a deterministic heuristic polynomial-time algorithm. So if one had faith in the analog of their approach, as applied to SAT, one would have to believe that under essentially every natural distribution SAT is easy (in the sense that there is an algorithm with a high probability weight of correctness under that distribution). Since the latter is not widely believed, we suggest that the right conclusion to draw from the main result of this section is simply that if one were to hope to effectively use on typical NP-complete sets the notion of juntas and of heuristic polynomial time w.r.t. juntas, one would almost certainly have to go beyond the basic three conditions and add additional conditions. Again, we stress that Procaccia and Rosenschein didn't focus on examples this far afield, and even within the world of mechanisms implied that unspecified additional conditions beyond the core three might be needed when studying problems other than voting manipulation problems. This section's contribution is to give a construction indicating that the core three junta conditions, standing on their own, seem too weak.

Since we will use the Procaccia–Rosenschein junta notion in a more general setting than merely manipulation problems, we to avoid any chance of confusion will use the term "basic junta" to denote that we have removed the word "manipulation" and that we are using their three "basic" properties, and not the two additional properties that are specific to voting manipulation. Our definition of "deterministic heuristic polynomial-time algorithm" is the same as theirs, and our definition of "basic deterministic heuristic polynomial-time algorithm" is the same as their notion of "susceptible" (we avoid the word "susceptible" as that term already has term-of-art meanings in the study of the complexity of elections, e.g., in [BTT92] and the line of work it started) except we have replaced the word "junta" with "basic junta"—and so again we are allowing their notion to be extended beyond just manipulation and mechanism problems.

**Definition 3.1**   1. *(see and contrast [PR07b]) Let $\mu = \{\mu_n\}_{n \in \mathbb{N}}$ be a distribution over the possible instances of an NP-hard problem L. (In this model, each $\mu_n$ sums to 1* *over all length n instances.[2]*) *We say $\mu$ is a* basic junta distribution *if and only if $\mu$ has the following properties:*

(a) **Hardness:** *The restriction of $L$ to $\mu$ is the problem whose possible instances are only $\bigcup_{n \in \mathbb{N}} \{x \mid |x| = n \text{ and } \mu_n(x) > 0\}$. Deciding this restricted problem is still NP-hard.*

(b) **Balance:** *There exist constants $c > 1$ and $N \in \mathbb{N}$ such that for all $n \geq N$ and for all instances $x$, $|x| = n$, we have $1/c \leq \text{Prob}_{\mu_n}[x \in L] \leq 1 - 1/c$.*

(c) **Dichotomy:** *There exists some polynomial $p$ such that for all $n$ and for all instances $x$, $|x| = n$, either $\mu_n(x) \geq 2^{-p(n)}$ or $\mu_n(x) = 0$.*

2. (see [PR07b]) *Let $(L, \mu)$ be a distributional decision problem (see Definition B.1 in Appendix B). An algorithm $\mathcal{A}$ is said to be a* deterministic heuristic polynomial-time *algorithm for $(L, \mu)$ if $\mathcal{A}$ is a deterministic polynomial-time algorithm and there exist a polynomial $q$ of degree at least one[3] and an $N \in \mathbb{N}$ such that for each $n \geq N$,*

$$\text{Prob}_{\mu_n}[x \notin L \iff \mathcal{A} \text{ accepts } x] < \frac{1}{q(n)}.$$

*When such a $\mu$ and $\mathcal{A}$ exist, we'll say that $L$ is in deterministic heuristic polynomial time (with respect to $\mu$).*

3. (see and contrast [PR07b]'s "susceptible") *Let $(L, \mu)$ be a distributional decision problem. An algorithm $\mathcal{A}$ is said to be a* basic deterministic heuristic polynomial-time *algorithm for $(L, \mu)$ if $\mu$ is a basic junta distribution (for $L$), $\mathcal{A}$ is a deterministic polynomial-time algorithm, and there exist a polynomial $q$ of degree at least 1 and $N \in \mathbb{N}$ such that for each $n \geq N$,*

$$\text{Prob}_{\mu_n}[x \notin L \iff \mathcal{A} \text{ accepts } x] < \frac{1}{q(n)}.$$

---

[2]The Procaccia–Rosenschein work clearly means this, as they state explicitly (see page 162 of their paper) that each $\mu_n$ is a distribution, and in addition all their work and notions are based on looking at a single length at a time and in their example of building a junta they (naturally, given what their model is) do not address relative weights between different lengths (and so a global distribution, i.e., one over $\Sigma^*$, is not being defined). (We mention in passing that if one were to try to interpret the notion as saying that there is a (Levin-like) single distribution over all lengths, one would in their definition of junta have foundational problems when that single distribution put no weight on any strings of a given length, as one would be faced with conditioning on a set of probability weight zero, which is not well-defined.) To avoid any confusion, we mention that although Procaccia–Rosenschein use the phrase "distributional problem," a term that as it is most typically used in the literature means the distribution is global, it is very clear that in their paper's clear and internally self-consistent nomenclature/definition framework when the words "distribution problem" are used the "distribution" is a collection of distributions, one per length. We mention in passing that our main theorem of this section, Theorem 3.3, remains true—though one has to shift the values in its proof a bit—even under the different case of having one global distribution. On the other hand, results such as nonclosure under polynomial-time isomorphisms potentially might not hold under that alternate model.

[3]Requiring that there exists a $q$ of degree at least 1, by which they of course mean nontrivially so, is basically ensuring that there exists a $c$ such that, in the displayed inequality of this definition, "$< c/n$" holds.

> *When such a $\mu$ and $\mathcal{A}$ exist, we'll say that $L$ is in basic deterministic heuristic polynomial time (with respect to $\mu$).*

We now explore their notion of deterministic heuristic polynomial time[4] and their notion of junta, both however viewed for general NP problems and using the "basic" three conditions. We will note that the notion in such a setting is in some senses not restrictive enough and in other senses is too restrictive. Let us start with the former. We need a definition.

**Definition 3.2** *We will say that a set $L$ is* well-pierced *(respectively,* uniquely well-pierced*) if there exist sets $Pos \in \mathrm{P}$ and $Neg \in \mathrm{P}$ such that $Pos \subseteq L$, $Neg \subseteq \overline{L}$, and there is some $N \in \mathbb{N}$ such that at each length $n \geq N$, each of Pos and Neg has at least one string at length $n$ (respectively, each of Pos and Neg has exactly one string at length $n$).*

Each uniquely well-pierced set is well-pierced. Note that, under quite natural encodings, such NP-complete sets as, for example, SAT certainly are well-pierced and uniquely well-pierced. (All this says is that, except for a finite number of exceptional lengths, there is one special string at each length that can easily, uniformly be recognized as in the set and one that can easily, uniformly be recognized as not in the set.) Indeed, under quite natural encodings, undecidable problems such as the halting problem are uniquely well-pierced.

Recall that juntas are defined in relation to an infinite list of distributions, one per length (so $\mu = \{\mu_n\}_{n \in \mathbb{N}}$). The Procaccia and Rosenschein definition of junta does not explicitly put computability or uniformity requirements on such distributions in the definition of junta, but it is useful to be able to make claims about that. So let us say that such a distribution is *uniformly computable in polynomial time* (respectively, is *uniformly computable in exponential time*) if there is a polynomial-time function (respectively, an exponential-time function) $f$ such that for each $i$ and each $x$, $f(i, x)$ outputs the value of $\mu_i(x)$ (say, as a rational number—if a distribution takes on other values, it simply will not be able to satisfy the notion of good uniform time).

**Theorem 3.3** *Let $A$ be any NP-hard set that is well-pierced. Then there exists a basic junta distribution relative to which $A$ has a basic deterministic heuristic polynomial-time algorithm (indeed, it even has a basic deterministic heuristic polynomial-time algorithm whose error weight is bounded not merely by $1/\mathrm{poly}$ (for some polynomial of degree at least 1) as the definition requires, but is even bounded by $1/2^{n^2-n}$). Furthermore, the junta is uniformly computable in exponential time, and if we in addition assume that $A$ is uniquely well-pierced, the junta is uniformly computable in polynomial time.*

---

[4]They credit their notion as being "inspired by Trevisan [Tre02] (there the same name is used for a somewhat different definition)." We mention in passing as an even earlier source for the same name, though also attached to a different definition than that of Procaccia and Rosenschein, Section 3 of [Imp95]. The "somewhat different" definitions aren't always so trivially different. For example, [BT06] gives an example that it says "seems absurd to consider [it] an efficient-on-average algorithm." The example does not meet [BT06]'s definition of heuristic polynomial time but does (barely—right at the border) satisfy the [PR07b] definition of heuristic polynomial time.

It follows that, under quite natural encodings, almost any natural set is in basic deterministic heuristic polynomial time. For example, SAT is and the halting problem is, both under natural encodings.[5] All it takes is for the given set to have at all but a finite number of lengths at least one element each that are uniformly easily recognizable as being in and out of the set.

**Proof.** Let $A$ be well-pierced. So there exists an $N$, and sets $Pos$ and $Neg$, that satisfy the definition of well-pierced. For each $n \geq N$, let $Pos(n)$ denote the lexicographically smallest length $n$ string in $Pos$ and let $Neg(n)$ denote the lexicographically smallest length $n$ string in $Neg$.

Define the distribution $\nu = \{\nu_n\}_{n \in \mathbb{N}}$ as follows:

1. For each length $n \geq N$, put weight $1/2^{n^2}$ on all length $n$ strings other than $Pos(n)$ and $Neg(n)$, and put weight $\frac{1}{2}\left(1 - \frac{2^n - 2}{2^{n^2}}\right)$ on each of $Pos(n)$ and $Neg(n)$.

2. For each length $n < N$, let $\nu_n$ be the uniform distribution over that length, i.e., each length $n$ string has weight $1/2^n$.

We now show that $\nu$ is a basic junta distribution.

1. Hardness: Since $\bigcup_{n \in \mathbb{N}} \{x \mid |x| = n \text{ and } \nu_n(x) > 0\}$ equals $\Sigma^*$, the restriction of $A$ to $\nu$ equals $A$, and so is still NP-hard.

2. Balance: Since for each length $n \geq N$ both $Pos(n) \in A$ and $Neg(n) \notin A$ have almost half of the probability weight of all length $n$ strings (namely, each has $\frac{1}{2}\left(1 - \frac{2^n - 2}{2^{n^2}}\right)$), $\nu$ is balanced.

3. Dichotomy: Since for all $n \geq N$ and for all $x$, $|x| = n$, we have $\nu_n(x) \geq 2^{-n^2}$, and for all $n < N$ and for all $x$, $|x| = n$, we have $\nu_n(x) \geq 2^{-n}$, dichotomy is satisfied.

Note that the junta is uniformly computable in exponential time, and if $A$ is uniquely well-pierced then the junta is uniformly computable in polynomial time.

Our basic deterministic heuristic polynomial-time algorithm for $(A, \nu)$ works as follows: On inputs that are a $Pos(n)$, it accepts; on inputs that are a $Neg(n)$, it rejects; and on every other input, it (for specificity, though it does not matter) accepts.

For each $n \geq N$, the error probability of this algorithm on inputs of length $n$ is at most $(2^n - 2)/2^{n^2} \leq 1/2^{n^2 - n}$. ❑

---

[5] Again, a potential problem when dealing with such claims is details of encoding. For example, if SAT is encoded in such a way that the vast majority of the strings (say, all but at most a $1/n$ portion of the strings) of each length are obviously syntactically illegal (and such encodings can indeed be totally natural), then an astute reader might well ask, "Isn't any algorithm that accepts the empty set a basic deterministic heuristic polynomial-time algorithm for SAT, relative to the uniform distribution, which obviously is a basic junta." However, this reasoning is flawed. If that many strings of each length are obviously syntactically illegal and we are using the uniform distribution, then the balance condition for juntas is violated. So the balance condition blocks that argument, and indeed this type of blocking is *precisely* why Procaccia and Rosenschein [PR07b] have the balance condition.

In the proof we achieve the error bound $1/2^{n^2-n}$ stated in Theorem 3.3. However, this bound can easily be strengthened to $1/2^{n^k-n}$, for each fixed constant $k$, by altering the proof. Note that the altered algorithm will depend on $k$.

Our point is not that this construction is difficult. Rather, our point is that this construction indicates that the basic three junta conditions on their own can be short-circuited, and thus a stronger set of conditions would be needed to seek to create a Procaccia–Rosenschein-type program against, e.g., SAT. More generally, one should probably be exceedingly skeptical about any distribution or distribution type that is being proposed—without proof—as perhaps being so hard that it seeks to "convincingly represent all other distributions with respect to average-case analysis" [PR07b, p. 163]. Again, we should stress that Procaccia and Rosenschein are clear that this is a hope rather than a claim, that they repeatedly stress that the approach may be controversial, and that their focus is on manipulation/mechanism issues.

Loosely put, the above result says that the basic junta conditions are in some ways overinclusive. We also note that the definition of junta, and the issue of when we will have a basic deterministic heuristic polynomial-time algorithm, are exceedingly sensitive to details of encoding.[6] We mention quickly two such effects, one that indirectly suggests overinclusiveness and one that suggests underinclusiveness.

As to the former, note that *every* NP-hard set is $\leq_m^p$-reducible to a set that is in basic deterministic heuristic polynomial time. This applies even to undecidable NP-hard sets, such as $\text{SAT} \oplus \text{HP} =_{def} \{0x \mid x \in \text{SAT}\} \cup \{1y \mid y \in \text{HP}\}$, where HP denotes the halting problem. The proof is nearly immediate. Given an NP-hard set $A$ (over some alphabet $\Sigma$ that has cardinality at least two, and w.l.o.g. we assume that 0 and 1 are letters of $\Sigma$), note that $A \leq_m^p$-reduces to the set $A' = \{00x \mid x \in \Sigma^*\} \cup \{1x1^{|x|^2+2} \mid x \in A\}$, and that $A'$ is easily seen to be in basic deterministic heuristic polynomial time (indeed, with error bound not just $1/poly$ but even $1/exponential$), in particular via the basic junta (relative to $A'$) that is the uniform distribution.

Regarding underinclusiveness, note that under the definition of basic junta, no set that at an infinite number of lengths either has all strings or has no strings can have a basic deterministic heuristic polynomial-time algorithm, since for such sets the balance condition of the notion of a basic junta can never be satisfied. It follows easily that the notion of having a basic deterministic heuristic polynomial-time algorithm is not even closed under polynomial-time isomorphisms.[7]

---

[6]In contrast, the "$\epsilon$" exponent and $|x|$ denominator (see Definition B.1 in Appendix B) in Levin's [Lev86] theory of AvgP, average-case polynomial-time, were precisely designed, in that different setting, to avoid such problems—problems that one gets by following the type of asymptotic focus on one length at a time that the Procaccia and Rosenschein model adopts. On the other hand, even Levin's theory has many subtleties and downsides, and to this day has not found anything resembling the type of widespread applicability of NP-completeness theory; see any of the many surveys on that topic.

[7]To be extremely concrete, the NP-complete set $B = \{00x \mid x \in \Sigma^*\} \cup \{1x1^{|x|^2+2} \mid x \in \text{SAT}\}$ is (as per the above) easily seen to be in basic deterministic heuristic polynomial time, but the NP-complete set $B' = \{xx \mid x \in \text{SAT}\}$, though it is by standard techniques polynomial-time isomorphic to $B$ (see [BH77]), is not in basic deterministic heuristic polynomial time. If the reader wonders why we did not simply use two

Finally, we mention in passing that it is important not to confuse the notion of having heuristic polynomial-time algorithms with the notion of having the average of one's time complexity be good. In particular, the former doesn't seem to imply the latter. Trevisan [Tre02] has already noted this effect, though under a definition of heuristic polynomial time that differs somewhat from that of Procaccia and Rosenschein [PR07b] (in particular differing precisely at the most interesting boundary, namely the $1/poly$). For completeness, we include as Appendix C an extended discussion of that effect (noted earlier by Trevisan under a slightly different definition and setting, but the point is the same).

## 4 Conclusions and Open Directions

We studied relationships between average-case polynomial time, benign algorithm schemes, and frequency (and probability weight) of correctness. We showed that all problems having benign algorithm schemes relative to the uniform distribution (and thus all sets in average-case polynomial time relative to the uniform distribution) have frequently self-knowingly correct algorithms.

We also studied, when limited to the "basic" three junta conditions, the notion of junta distributions and of basic deterministic heuristic polynomial time, and we showed that they admit some extreme behaviors. It might be natural for the reader to ask us to provide a set of alternate conditions to define some altered notion of junta to achieve the goal that all distributions meeting the conditions focus so much weight on "hard" instances that (loosely put) if a problem is easy relative to a junta then it will be easy relative to any reasonable distribution (such as the uniform distribution). However, although we think this is a lovely framework and would be very happy indeed if some set of conditions as natural, broad, and simple as those of Procaccia and Rosenschein [PR07b] will achieve this, we simply do not believe that that is the case. Our intuition is that simple, broad, transparent conditions won't be enough to filter out any broad class of simple distributions each of which is as challenging as possible to all relevant problems.

However, as encouragement to those who hope otherwise, we mention the relevant (as one could succeed simply by making one's junta condition be that the distribution must be the simply defined universal distribution discussed in Li–Vitányi, if one felt that focusing on that nonrecursive distribution was a fair approach) and stunning result of Li and Vitányi ([LV92], and for a more extensive history, e.g., regarding the Solomonoff–Levin measure, see also [Mil93]) who showed that a certain simply described universal distribution (unfortunately, one that is so complex as to itself be nonrecursive) based on Kolmogorov complexity has the property that it causes the average-case and worst-case complexities (of algorithms) to coincide.

As a concrete suggestion to those who would like to adjust starting from the current basic three conditions, as a possible modification we would suggest removing the Dichotomy Condition (which as one can see in the work of the present paper can be satisfied too

---

P sets, the reason is, under the Procaccia–Rosenschein definition, one needs NP-hardness to have a junta, and one needs a junta to put something in deterministic heuristic polynomial time.

easily) and replacing it with an "Almost-Uniformity Condition," which we define to be the requirement that there be a positive constant $K$ and a natural number $n_0$ such that for each $n > n_0$ and for all $x$ and $y$ satisfying $|x| = |y| = n$, it holds that if $\mu_n(x) \neq 0$ and $\mu_n(y) \neq 0$ then $\mu_n(x)/\mu_n(y) \leq K$. That is, all strings at each (except excluding perhaps a finite set of lengths) given length that get more than zero weight don't differ from each other by more than a (global for the distribution) multiplicative constant. The dichotomy condition does not itself ensure almost-uniformity. Indeed, our proof of Theorem 3.3 badly violates almost-uniformity, and so replacing Dichotomy with Almost-Uniformity at least invalidates that particular proof. Although the focus of this paper has not been on the five-condition suite of conditions that include two additional ones tailored to the needs of NP-hard voting manipulation problems, we mention the interesting fact that, as pointed out to us by Ariel Procaccia [Pro07], the manipulation-related distribution defined in [PR07b, Section 4.1] and used powerfully in their paper not only satisfies their five conditions but also satisfies almost-uniformity.

**Acknowledgments:** Earlier versions of parts of this paper appeared in COMSOC 2006 and FCT 2007. We are deeply grateful to anonymous referees and Ariel Procaccia for their helpful comments on precursors of and earlier versions of this paper, to Tuomas Sandholm for the very important comments mentioned in the footnote to the paper's title, to Chris Homan for his interest in this work and for many inspiring discussions on computational issues related to voting, and to Osamu Watanabe for hosting a visit during which this work was done in part. As always, any errors and all opinions are the sole responsibility of the authors.

# References

[BH77]   L. Berman and J. Hartmanis. On isomorphisms and density of NP and other complete sets. *SIAM Journal on Computing*, 6(2):305–322, 1977.

[Bla58]   D. Black. *The Theory of Committees and Elections*. Cambridge University Press, 1958.

[BT06]   A. Bogdanov and L. Trevisan. *Average-Case Complexity*. Now Publishers, 2006.

[BTT89]   J. Bartholdi III, C. Tovey, and M. Trick. Voting schemes for which it can be difficult to tell who won the election. *Social Choice and Welfare*, 6(2):157–165, 1989.

[BTT92]   J. Bartholdi III, C. Tovey, and M. Trick. How hard is it to control an election? *Mathematical Comput. Modelling*, 16(8/9):27–40, 1992.

[CS06]   V. Conitzer and T. Sandholm. Nonexistence of voting rules that are usually hard to manipulate. In *Proceedings of the 21st National Conference on Artificial Intelligence*. AAAI Press, July 2006.

[Dod76]    C. Dodgson. A method of taking votes on more than two issues. Pamphlet printed by the Clarendon Press, Oxford, and headed "not yet published" (see the discussions in [MU95, Bla58], both of which reprint this paper), 1876.

[EHRS07a]  G. Erdélyi, L. Hemaspaandra, J. Rothe, and H. Spakowski. On approximating optimal weighted lobbying, and frequency of correctness versus average-case polynomial time. In *Proceedings of the 16th International Symposium on Fundamentals of Computation Theory*, pages 300–311. Springer-Verlag *Lecture Notes in Computer Science #4639*, August 2007.

[EHRS07b]  G. Erdélyi, L. Hemaspaandra, J. Rothe, and H. Spakowski. On approximating optimal weighted lobbying, and frequency of correctness versus average-case polynomial time. Technical Report TR-914, Department of Computer Science, University of Rochester, Rochester, NY, March 2007.

[Fei02]    U. Feige. Relations between average-case complexity and approximation complexity. In *Proceedings of the 34th ACM Symposium on Theory of Computing*, pages 534–543. ACM Press, May 2002.

[FHHR]     P. Faliszewski, E. Hemaspaandra, L. Hemaspaandra, and J. Rothe. A richer understanding of the complexity of election systems. In S. Ravi and S. Shukla, editors, *Fundamental Problems in Computing: Essays in Honor of Professor Daniel J. Rosenkrantz*. Springer. To appear. Available as Technical Report cs.GT/0609112, ACM Computing Research Repository (CoRR), September 2006.

[Fis77]    P. Fishburn. Condorcet social choice functions. *SIAM Journal on Applied Mathematics*, 33(3):469–489, 1977.

[FKN07]    E. Friedgut, G. Kalai, and N. Nisan. Elections can be manipulated often. Manuscript, URL cs.huji.ac.il/∼noam, 2007.

[Gol97]    O. Goldreich. Notes on Levin's theory of average-case complexity. Technical Report TR97-058, Electronic Colloquium on Computational Complexity, November 1997.

[HH]       C. Homan and L. Hemaspaandra. Guarantees for the success frequency of an algorithm for finding Dodgson-election winners. *Journal of Heuristics*. To appear. Conference version available as [HH06] and full version available as [HH05].

[HH05]     C. Homan and L. Hemaspaandra. Guarantees for the success frequency of an algorithm for finding Dodgson-election winners. Technical Report TR-881, Department of Computer Science, University of Rochester, Rochester, NY, September 2005. Revised, June 2007.

[HH06]       C. Homan and L. Hemaspaandra. Guarantees for the success frequency of an algorithm for finding Dodgson-election winners. In *Proceedings of the 31st International Symposium on Mathematical Foundations of Computer Science*, pages 528–539. Springer-Verlag *Lecture Notes in Computer Science #4162*, August/September 2006.

[HHR97]     E. Hemaspaandra, L. Hemaspaandra, and J. Rothe. Exact analysis of Dodgson elections: Lewis Carroll's 1876 voting system is complete for parallel access to NP. *Journal of the ACM*, 44(6):806–825, November 1997.

[Imp95]     R. Impagliazzo. A personal view of average-case complexity. In *Proceedings of the 10th Structure in Complexity Theory Conference*, pages 134–147. IEEE Computer Society Press, 1995.

[Lev86]     L. Levin. Average case complete problems. *SIAM Journal on Computing*, 15(1):285–286, 1986.

[LV92]      M. Li and P. Vitányi. Average case complexity under the universal distribution equals worst-case complexity. *Information Processing Letters*, 42(3):145–149, 1992.

[Mil93]     P. Miltersen. The complexity of malign measures. *SIAM Journal on Computing*, 22(1):147–156, 1993.

[MPS]       J. McCabe-Dansted, G. Pritchard, and A. Slinko. Approximability of Dodgson's rule. *Social Choice and Welfare*. To appear. Conference version available as [MPS06].

[MPS06]     J. McCabe-Dansted, G. Pritchard, and A. Slinko. Approximability of Dodgson's rule. In U. Endriss and J. Lang, editors, *Proceedings of the 1st International Workshop on Computational Social Choice*, pages 331–344. Universiteit van Amsterdam, December 2006. Available online at staff.science.uva.nl/∼ulle/COMSOC-2006/proceedings.html.

[MU95]      I. McLean and A. Urken. *Classics of Social Choice*. University of Michigan Press, Ann Arbor, Michigan, 1995.

[PR07a]     A. Procaccia and J. Rosenschein. Average-case tractability of manipulation in voting via the fraction of manipulators. In *Proceedings of the 6th International Joint Conference on Autonomous Agents and Multiagent Systems*, pages 718–720, May 2007.

[PR07b]     A. Procaccia and J. Rosenschein. Junta distributions and the average-case complexity of manipulating elections. *Journal of Artificial Intelligence Research*, 28:157–181, 2007.

[Pro07]     A. Procaccia, April 8, 2007. Personal communication.

[RSV03]  J. Rothe, H. Spakowski, and J. Vogel. Exact complexity of the winner problem for Young elections. *Theory of Computing Systems*, 36(4):375–386, June 2003.

[SSGL02]  T. Sandholm, S. Suri, A. Gilpin, and D. Levine. Winner determination in combinatorial auction generalizations. In *Proceedings of the 1st International Joint Conference on Autonomous Agents and Multiagent Systems*, pages 69–76. ACM Press, July 2002.

[Tre02]  L. Trevisan. Lecture notes on computational complexity. www.cs.berkeley.edu/~luca/notes/complexitynotes02.pdf (Lecture 12), 2002.

[Wan97]  J. Wang. Average-case computational complexity theory. In L. Hemaspaandra and A. Selman, editors, *Complexity Theory Retrospective II*, pages 295–328. Springer-Verlag, 1997.

## A  Homan and Hemaspaandra's Frequently Self-Knowingly Correct Greedy Algorithm

Homan and Hemaspaandra [HH] proposed the following definition of a new type of algorithm to capture the notion of "guaranteed high success frequency" formally.

**Definition A.1 ([HH])**   *1. Let $f : S \to T$ be a function, where $S$ and $T$ are sets. We say an algorithm $\mathcal{A} : S \to T \times \{$ "definitely", "maybe"$\}$ is self-knowingly correct for $f$ if, for each $s \in S$ and $t \in T$, whenever $\mathcal{A}$ on input $s$ outputs $(t,$ "definitely"$)$ then $f(s) = t$.*

  2. *An algorithm $\mathcal{A}$ that is self-knowingly correct for $g : \Sigma^* \to T$ is said to be* frequently self-knowingly correct for $g$ if

$$\lim_{n \to \infty} \frac{\|\{x \in \Sigma^n \mid A(x) \in T \times \{\text{``maybe''}\}\}\|}{\|\Sigma^n\|} = 0.$$

In their paper [HH], Homan and Hemaspaandra present two frequently self-knowingly correct polynomial-time algorithms, which they call Greedy-Score and Greedy-Winner. Since Greedy-Winner can easily be reduced to Greedy-Score, we focus on Greedy-Score only and briefly describe the intuition behind this algorithm; for full detail, we refer to [HH]. (But both heuristics work well tremendously often—in a formal sense of the notion—provided that the number of voters greatly exceeds the number of candidates.)

If $(C, V)$ is an election and $c$ is some designated candidate in $C$, we call $(C, V, c)$ a *Dodgson triple*. Given a Dodgson triple $(C, V, c)$, Greedy-Score determines the Dodgson score of $c$ with respect to the given election $(C, V)$. We will see that there are Dodgson triples $(C, V, c)$ for which this problem is particularly easy to solve.

For any $d \in C - \{c\}$, let Deficit$[d]$ be the number of votes $c$ needs to gain in order to have more votes than $d$ in a pairwise contest between $c$ and $d$.

**Definition A.2** *Any Dodgson triple $(C, V, c)$ is said to be* nice *if for each candidate $d \in C - \{c\}$, there are at least $\text{Deficit}[d]$ votes for which candidate $c$ is exactly one position below candidate $d$.*

Given a Dodgson triple $(C, V, c)$, the algorithm Greedy-Score works as follows:

1. For each candidate $d \in C - \{c\}$, determine $\text{Deficit}[d]$.

2. If $(C, V, c)$ is not nice then output ("anything","maybe"); otherwise, output

$$(\sum_{d \in C - \{c\}} \text{Deficit}[d], \text{``definitely''}).$$

Note that, for nice Dodgson triples, we have

$$\text{DodgsonScore}(C, V, c) = \sum_{d \in C - \{c\}} \text{Deficit}[d],$$

It is easy to see that Greedy-Score is a self-knowingly correct polynomial-time bounded algorithm. To show that it is even *frequently* self-knowingly correct, Homan and Hemaspaandra prove the following lemma. Their proof uses a variant of Chernoff bounds.

**Lemma A.3 (Thm. 4.1.3 of [HH])** *Let $(C, V, c)$ be a given Dodgson triple with $n = \|V\|$ votes and $m = \|C\|$ candidates, chosen uniformly at random among all such Dodgson elections. The probability that $(C, V, c)$ is not nice is at most $2(m-1)e^{-\frac{n}{8m^2}}$.*

Homan and Hemaspaandra [HH] show that the heuristic Greedy-Winner, which is based on Greedy-Score and which solves the winner problem for Dodgson elections, also is a frequently self-knowingly correct polynomial-time algorithm. This result is stated formally below.

**Theorem A.4 (Thm. 4.4.2 of [HH])** *For all $m, n \in \mathbb{N}^+$, the probability that a Dodgson election $(C, V)$ selected uniformly at random from all Dodgson elections having $m$ candidates and $n$ votes (i.e., all $(m!)^n$ Dodgson elections having $m$ candidates and $n$ votes have the same likelihood of being selected) has the property that there exists at least one candidate $c$ such that Greedy-Winner on input $(C, V, c)$ outputs "maybe" as its second output component is less than $2(m^2 - m)e^{-\frac{n}{8m^2}}$.*

## B  Foundations of Average-Case Complexity Theory

The theory of average-case complexity was initiated by Levin [Lev86]. A problem's average-case complexity can be viewed as a more significant measure than its worst-case complexity in many cases, for example in cryptographic applications. We here follow Goldreich's presentation [Gol97]. Another excellent introduction to this theory is that of Wang [Wan97].

Fix the alphabet $\Sigma = \{0, 1\}$, let $\Sigma^*$ denote the set of strings over $\Sigma$, and let $\Sigma^n$ denote the set of all length $n$ strings in $\Sigma^*$. For any $x, y \in \Sigma^*$, $x < y$ means that $x$ precedes $y$ in lexicographic order, and $x - 1$ denotes the lexicographic predecessor of $x$.

Intuitively, Levin observed that many hard problems—including those that are NP-hard in the traditional worst-case complexity model—might nonetheless be easy to solve "on the average," i.e., for "most" inputs or for "most practically relevant" inputs. He proposed to define the complexity of problems with respect to some suitable distribution on the input strings.

We now define the notion of a distributional problem and the complexity class AvgP.

Here, we define only distributional search problems; the definition of distributional decision problems is analogous.

**Definition B.1 ([Lev86], see also [Gol97, Wan97])** *1. A distribution function $\mu : \Sigma^* \to [0, 1]$ is a nondecreasing function from strings to the unit interval that converges to one (i.e., $\mu(0) \geq 0$, $\mu(x) \leq \mu(y)$ for each $x < y$, and $\lim_{x \to \infty} \mu(x) = 1$). The density function associated with $\mu$ is defined by $\mu'(0) = \mu(0)$ and $\mu'(x) = \mu(x) - \mu(x-1)$ for each $x > 0$. That is, each string $x$ gets weight $\mu'(x)$ with this distribution.*

*2. A distributional (search) problem is a pair $(f, \mu)$, where $f : \Sigma^* \to \Sigma^*$ is a function and $\mu : \Sigma^* \to [0, 1]$ is a distribution function.*

*3. A function $t : \Sigma^* \to \mathbb{N}$ is polynomial on the average with respect to some distribution $\mu$ if there exists a constant $\epsilon > 0$ such that*

$$\sum_{x \in \Sigma^*} \mu'(x) \cdot \frac{t(x)^\epsilon}{|x|} < \infty.$$

*4. The class AvgP consists of all distributional problems $(f, \mu)$ for which there exists an algorithm $\mathcal{A}$ computing $f$ such that the running time of $\mathcal{A}$ is polynomial on the average with respect to the distribution $\mu$.*

In Section 2.2, we focused on the standard uniform distribution $\hat{\mu}$ on $\Sigma^*$, which is defined by

$$\hat{\mu}'(x) = \frac{1}{|x|(|x| + 1)2^{|x|}}.$$

That is, we first choose an input size $n$ at random with probability $1/(n(n + 1))$, and then we choose an input string of that size $n$ uniformly at random.

In Section 2.2, we considered polynomial-time benign algorithm schemes. This notion was introduced by Impagliazzo [Imp95] to provide an alternative view on the definition of Levin's class AvgP (average polynomial time, see [Lev86]).

We in this paper use the following notation. For any distribution $\mu$ and for each $n \in \mathbb{N}$ let $\mu_n$ be the restriction of $\mu$ to strings of length exactly $n$, and let $\mu_{\leq n}$ be the restriction of $\mu$ to strings of length at most $n$. (When discussing benign algorithms, the length-0 string $\epsilon$ is routinely completely excluded from the probability distribution—it is by convention given weight zero—and so for such cases, e.g., Definition B.2, the "for each $n \in \mathbb{N}$" should be viewed as changed to saying "for each $n \in \mathbb{N}^+$.")

**Definition B.2 ([Imp95])**   1. *An algorithm computes a function $f$ with* benign faults *if it either outputs an element of the image of $f$ or "?," and if it outputs anything other than ?, it is correct.*

2. *Let $\mu$ be a distribution on $\Sigma^*$. A* polynomial-time benign algorithm scheme *for a function $f$ on $\mu$ is an algorithm $\mathcal{A}(x, \delta)$ such that:*

    (a) *$\mathcal{A}$ runs in time polynomial in $|x|$ and $1/\delta$.*
    
    (b) *$\mathcal{A}$ computes $f$ with benign faults.*
    
    (c) *For each $\delta$, $0 < \delta < 1$, and for each $n \in \mathbb{N}^+$,*
    
    $$\mathrm{Prob}_{\mu_{\leq n}}[\mathcal{A}(x, \delta) = \ ?] \leq \delta.$$

## C   Some Comments on Heuristic Polynomial-Time Versus Average-Case Running Times

Procaccia and Rosenschein [PR07b] in the title and body of their paper describe their theory as being an "average-case complexity" theory for manipulation. In fact, as their definitions then carefully make clear, they are not using the term in any way that involves taking an average of running times, but rather their theory is actually an approach to adding probability weights to a frequency of correctness approach. Nonetheless, the use of the term makes it natural for the reader to wonder whether something being simple in their unusual use of "average-case" implies its average running-time is also low. (As to how usual or unusual their use is, one actually can find a small but perhaps growing number of cases in the complexity literature (see [Fei02, BT06]) of the term "average case" being used not in a direct averaging sense and not in the sense of Levin's transformed averaging-like theory [Lev86] but rather for something regarding frequency of correctness, although though even in such cases authors typically use a term such as "average-case tractable," and carefully avoid using a term such as "average(-case) *polynomial*" to refer to frequency claims.)

So, in brief, Procaccia and Rosenschein [PR07b] look at probability weight (relative to some distribution) of correctness. However, they require correctness only for $1 - 1/poly$ probability weight at each length. This is quite a lot of probability weight, but it seems not anywhere near enough to ensure good average-case complexity. Let us consider taking a deterministic heuristic polynomial-time algorithm for a problem and trying to build from it an algorithm for the problem (i.e., one that is correct on all instances) that has good average-case running time. First, notice that one is dead from the start, as deterministic heuristic polynomial-time algorithms, though very frequently correct, are not guaranteed to know any easily recognized broad set of inputs on which they are guaranteed to be correct. But for the sake of argument, let us suppose that for our given deterministic heuristic polynomial-time algorithm we by good luck have that there is a deterministic polynomial-time set, having at each length probability weight under the junta at least $1 - 1/poly$, such that for each element of this set the deterministic heuristic polynomial-time algorithm is correct. Of course, for the remaining $1/poly$ of the weight the algorithm might be correct or not correct—no guarantees. Even with this strong extra assumption (which is basically tossing in a Homan–Hemaspaandra-like self-knowing correctness property, see [HH]), the average-case time analysis doesn't come out happily. Note that for the remaining $1/poly$ of the probability weight (and we are assuming that this is not just an upper bound but that one might actually have about this much weight for these bad cases), one would have to potentially brute force those, and as Procaccia and Rosenschein focus on NP-hard problems, each such brute-forcing would seem to potentially take exponential time. So, very loosely put, and under all the assumptions we are making (e.g., about having to use brute force and so on), the expected time (over their own distribution) one gets is roughly $\left(1 - \frac{1}{poly}\right) \cdot poly' + \frac{exponential}{poly}$. And, critically, that is exponential. (What we just argued is that, very informally, in the model of looking at junta-weighted average time over the strings of each length and looking at the asymptotics of that, the obvious attempt to convert a deterministic heuristic polynomial-time algorithm into an algorithm (i.e., a correct program for the problem) with good average-case running time yields an exponential average. However, it is true that such asymptotics of averages over each length have in other settings some undesirably properties, see, e.g., [Gol97]. Nonetheless, the $1/poly$ weight of the exponential time here is so bad that going into an even more Levin-like setting by tamping down on the runtimes by adding an "$\epsilon$" exponent still would not seem, if done naturally, to tame the exponential nature of the average time.)

So in summary Procaccia and Rosenschein's notion, which is not based on taking averages, is best viewed as a shift of the nature of "frequency of correctness" approaches to focus instead on "probability weight of correctness (relative to some distribution)"—which is a quite natural shift to look at.

We mention in passing that the papers of Homan and Hemaspaandra [HH] and of Conitzer and Sandholm [CS06] are about frequency of correctness—and are quite explicit that that, and not average-case complexity, is what they are about. Despite the fact that the work of Homan and Hemaspaandra [HH] ensures not just a $1/poly$-bounded weight of bad cases but indeed an at most $1/exponential$ proportion (note: it is in a uniform-like model), a (very informal) argument similar to the one above can be made suggesting that the work of Homan and Hemaspaandra also does not seem to in any obvious way yield an average-case polynomial-time claim; that argument can be found in an earlier version of this paper [EHRS07b].

Regarding the above comments of this appendix, please make sure to note Section 3's pointer to Trevisan's earlier, much-related observation of the core effect (although that is under a slightly different definition).

We commend to the attention of the reader interested in issues of frequency of success of manipulation the recent papers [PR07a, FKN07].

This ends our appendix on heuristic polynomial time versus average-case running times. Our general suggestion would be that, for clarity, when speaking of easiness the term "average-case" be reserved either for some direct averaging or for Levin's class AvgP, and that when speaking of frequency or probability weight of correctness, the term "average-case" be avoided and that instead terms such as "typical(-case) polynomial time" or "heuristic polynomial time" be used; see [HH] for additional discussion.